\documentclass[a4paper,11pt]{article}
\usepackage{pos}
\usepackage{xspace}
\usepackage{xcolor}

\makeatletter
\newcommand\myparagraph{\@startsection{paragraph}{4}{\z@}%
  {-5\p@ \@plus -3\p@ \@minus -3\p@}%
  {2\p@}%
  {\normalfont\itshape}%
}

\newcommand\mysection{\@startsection{section}{1}{\z@}%
                                   {2.5ex \@plus 1.0ex \@minus .5ex}%
                                   {1.5ex \@plus.3ex \@minus .3ex}%
                                   {\normalfont\large\secstyle}}
\makeatother

\newcommand{\MSbar}{\ensuremath{\overline{\text{MS}}}\xspace}
\newcommand{\ri}{\mathrm i}
\newcommand{\rw}{{\mathrm{w}}}

\newcommand{\ren}{{\mathrm{R}}}
\newcommand{\bare}{{\mathrm{B}}}

\newcommand{\FJTS}{\mathrm{FJTS}}
\newcommand{\PRTS}{\mathrm{PRTS}}
\newcommand{\GIVS}{\mathrm{GIVS}}

\newcommand{\OS}{\mathrm{OS}}
\newcommand{\EW}{{\mathrm{EW}}}

\newcommand{\GeV}{\ensuremath{\,\text{GeV}}\xspace}

\newcommand{\PH}{\ensuremath{\text{H}}\xspace}

\newcommand{\Ph}{\ensuremath{\text{h}}\xspace}
\newcommand{\PA}{\ensuremath{\text{A}}\xspace}

\newcommand{\Pb}{\ensuremath{\text{b}}\xspace}

\newcommand{\Pt}{\ensuremath{\text{t}}\xspace}

\newcommand{\PW}{\ensuremath{\text{W}}\xspace}
\newcommand{\PZ}{\ensuremath{\text{Z}}\xspace}

\newcommand{\PhiM}{\mathbf{\Phi}}

\newcommand{\bfzeta}{\text{\boldmath{$\zeta$}}}

\newcommand{\MH}{\ensuremath{M_\PH}\xspace}

\newcommand{\MW}{\ensuremath{M_\PW}\xspace}
\newcommand{\MZ}{\ensuremath{M_\PZ}\xspace}
\newcommand{\sw}{s_{\rw}}

\newcommand{\id}{{\rm 1\kern-.12em
 \rule{0.3pt}{1.5ex}\raisebox{0.0ex}{\rule{0.1em}{0.3pt}}}}

\title{Electroweak renormalization based on gauge-invariant vacuum expectation values}

\author*[a]{Stefan Dittmaier}
\author[b]{Heidi Rzehak}

\affiliation[a]{Physikalisches Institut, Albert-Ludwigs-Universit\"at Freiburg, \\
  Hermann-Herder-Str.~3, 79104 Freiburg, Germany, Country}

\affiliation[b]{Institute for Theoretical Physics, University of T\"ubingen,\\
Auf der Morgenstelle 14, 72076 T\"ubingen, Germany}

\emailAdd{stefan.dittmaier@physik.uni-freiburg.de}
\emailAdd{heidi.rzehak@itp.uni-tuebingen.de}

\abstract{We briefly review a recently proposed scheme for a gauge-invariant 
treatment of tadpole corrections in spontaneously broken gauge theories called 
{\it Gauge-Invariant Vacuum expectation value Scheme (GIVS)}.
The tadpole scheme matters in higher-order predictions of observables if
not all free parameters are fixed by renormalization conditions based on
S-matrix elements, such as in $\MSbar$ renormalization.
In contrast to previously used tadpole schemes, the GIVS unifies the 
properties of gauge invariance and perturbative stability. 
The application of the GIVS to the Standard Model, for instance, leads
to very moderate electroweak corrections in the conversion of
on-shell-renormalized to $\MSbar$-renormalized masses. 
Moreover,
in models with extended Higgs sectors, the GIVS is less prone to 
perturbative instabilities in the $\MSbar$ renormalization of Higgs
mixing angles than observed for the traditional gauge-independent 
tadpole treatment.
We illustrate this by considering the next-to-leading-order
(electroweak and QCD) corrections
to the decay processes $\Ph/\PH\to\PW\PW/\PZ\PZ\to4\,$fermions of the
CP-even neutral Higgs bosons~h and H
in a singlet Higgs extension of the Standard Model and
in the Two-Higgs-Doublet Model.

}

\FullConference{%
  Loops and Legs in Quantum Field Theory - LL2022,\\
  25-30 April, 2022\\
  Ettal, Germany
}

\begin{document}
\maketitle

\mysection{Introduction}

Electroweak (EW) corrections in the Standard Model (SM) and its extensions
are an important ingredient in precision calculations
for present and future collider phenomenology.
Renormalization, which is an important part of this task, in particular
includes the definition of vacuum expectation value (vev) parameters, 
such as $v$ in the SM, in higher orders---an issue that 
is connected to the treatment of tadpole loop contributions.
In case, all model parameters are fixed by renormalization 
conditions based on S-matrix elements, i.e.\ by so-called {\it on-shell (OS)}
conditions, higher-order predictions of observables do not depend on
the tadpole scheme. However, this is different, for instance, in $\MSbar$ renormalization schemes,
which are frequently employed for mass parameters and mixing angles in
extended Higgs sectors.

Explicit tadpole contributions are most conveniently removed via
tadpole coun\-ter\-terms which can be generated in two different ways in the Lagrangian:
in the course of parameter renormalization~\cite{Bohm:1986rj,Denner:1991kt}, 
or alternatively via Higgs field 
redefinitions~\cite{Fleischer:1980ub}.
The former, called {\it Parameter Renormalized Tadpole Scheme (PRTS)} in the following,
typically leads to small corrections originating from tadpoles, but
in general suffers from gauge dependences
if $\MSbar$ renormalization conditions are used for mass parameters.
The latter, usually called {\it Fleischer--Jegerlehner Tadpole Scheme (FJTS)}%
\footnote{The FJTS~\cite{Fleischer:1980ub} is equivalent to the $\beta_t$ scheme of
Ref.~\cite{Actis:2006ra}.},
is free from gauge dependences, but is prone to
very large corrections in $\MSbar$ schemes,
jeopardizing perturbative stability of $\MSbar$ predictions.
More details on tadpole renormalization can, e.g., be found
in Refs.~\cite{Dittmaier:2022maf,Dittmaier:2022ivi,Krause:2016oke,Denner:2016etu,%
Denner:2019vbn,Dudenas:2020ggt}.

In Refs.~\cite{Dittmaier:2022maf,Dittmaier:2022ivi} 
we have proposed a new scheme for tadpole renormalization,
dubbed {\it Gauge-Invariant Vacuum expectation value
Scheme (GIVS)}, which is a hybrid scheme
of the two mentioned types, with the benefits of being gauge independent 
and perturbatively stable.%
\footnote{Based on completely different considerations about symmetry breaking
in $R_\xi$ gauges, a similar idea for a hybrid scheme
has been mentioned in Sect.~D of Ref.~\cite{Dudenas:2020ggt}.}
The GIVS is based on the gauge-invariance property of Higgs fields, and the corresponding
parameters like $v$, in non-linear representations of Higgs multiplets.
In this article we briefly summarize the salient features of the GIVS
and its first applications within the SM~\cite{Dittmaier:2022maf}, 
a Singlet Higgs Extension of the SM (SESM)~\cite{Dittmaier:2022ivi}, and
to the Two-Higgs-Doublet Model (THDM)~\cite{Dittmaier:2022ivi}.

\mysection{Non-linear representations of Higgs sectors}

One of the basic ideas underlying the GIVS is the use of a field
basis in which the Higgs fields developing non-vanishing vevs are
gauge invariant. This implies that explicit tadpole contributions
induced by loop diagrams are gauge independent.
In such a field basis the would-be Goldstone-boson part of
the scalar Lagrangian is necessarily parametrized in a non-linear fashion. 
In the following, we briefly sketch the structure of
appropriate non-linear representations for the SM, the SESM, and the
THDM. For the full details, we refer to
Refs.~\cite{Dittmaier:2022maf,Dittmaier:2022ivi}.

\myparagraph{Standard Model}

Non-linear field representations are most conveniently formulated
via matrix fields. Denoting the usual, two-component SM Higgs doublet $\Phi$
and its charge conjugate $\Phi^{\mathrm{c}}$, we form the $2\times2$ matrix
$\PhiM =\left( \Phi^{\mathrm{c}},\Phi \right)$ and write it in some
``polar representation''
\begin{align}\label{eq:PhinonlinSM}
  \PhiM = {\textstyle\frac{1}{\sqrt{2}}} (v + h) U(\bfzeta), \qquad
U(\bfzeta) \equiv \exp \left(\frac{2\ri \bfzeta}{v}\right), \qquad
\bfzeta \equiv {\textstyle\frac{1}{2}}{\zeta_j\sigma_j},
\end{align}
in which $h$ corresponds to the physical Higgs field, which is gauge invariant, 
$\zeta_j$ ($j=1,2,3$) are real would-be Goldstone-boson fields
and $\sigma_j$ denotes the Pauli matrices. 
Since the matrix $U(\bfzeta)$ is unitary, the gauge-invariant square 
$\mathrm{tr}\bigl[\PhiM^\dagger\PhiM\bigr] = (v+h)^2$ does not
involve would-be Goldestone-boson fields $\zeta_j$.
In the SM Higgs Lagrangian
\begin{align}
{\cal L}_{\text{H,SM}} &=
{\textstyle\frac{1}{2}}\text{tr}\left[(D_\mu \PhiM)^\dagger(D^\mu \PhiM)\right] - V_{\text{SM}}
\end{align}
the first term is the kinetic part, which encodes the gauge interactions
of the scalar fields in the covariant derivative $D_\mu$.
Owing to the non-linear representation of $\PhiM$, this part is
non-polynomial in $\zeta_j$ and induces interaction vertices of arbitrarily many
would-be Goldstone-boson fields, but this is only a minor complication,
and the usual perturbative Feynman diagram calculus works as usual.
Note that the SM Higgs potential
\begin{align}
V_{\text{SM}} = -{\textstyle\frac{1}{2}} \mu_{2}^2 \mathrm{tr}\bigl[\PhiM^{\dag}\PhiM\bigr]
+{\textstyle\frac{1}{16}}{\lambda_2}\bigl(\mathrm{tr}\bigl[\PhiM^{\dag}\PhiM\bigr]\bigr)^2
= -{\textstyle\frac{1}{2}}\mu_{2}^2(v+h)^2 +{\textstyle\frac{1}{16}}{\lambda_2} (v+h)^4,
\end{align}
with the free parameters $\mu_{2}^2$ and $\lambda_2$,
is free from would-be Goldstone-boson fields.
With the of help the Nielsen identities~\cite{Gambino:1999ai}, it is easy to show
that the one-point vertex function $\Gamma^h_{\mathrm{nl}}$ of the
gauge-invariant Higgs field $h$ is gauge independent;
a simple one-loop calculation confirms this.

\myparagraph{Singlet extension of the SM (SESM)}

The SESM extends the SM by a real singlet scalar field $\sigma$, leading to a second
CP-even Higgs boson~H.
The SESM Higgs Lagrangian is given by
\begin{align}
{\cal L}_{\text{H,SESM}} &{}=
{\textstyle\frac{1}{2}}(\partial \sigma)^2+
{\textstyle\frac{1}{2}}\text{tr}\left[(D_\mu \PhiM)^\dagger(D^\mu \PhiM)\right]
- V_{\text{SESM}},
\end{align}
with a non-polynomial kinetic part similar as in the SM
and the gauge-invariant Higgs potential 
\begin{align}
V_{\text{SESM}} ={}& -{\textstyle\frac{1}{2}}{\mu_{2}^2} \mathrm{tr}\bigl[\PhiM^\dagger \PhiM\bigr]  - \mu_{1}^2 \sigma^2
  +{\textstyle\frac{1}{16}}{\lambda_{2}} \big(\mathrm{tr}\bigl[\PhiM^\dagger \PhiM\bigr]\big)^2
  +\lambda_{1} \sigma^4
  +{\textstyle\frac{1}{2}}{\lambda_{12}} \mathrm{tr}\bigl[\PhiM^\dagger \PhiM\bigr]\,\sigma^2,
\end{align}
which again does not involve would-be Goldstone-boson fields $\zeta_j$.
The Higgs fields $h_1$ and $h_2$
corresponding to particle excitations in $\sigma$ and $\PhiM$, respectively,
are identified by
\begin{align}\label{eq:PhinonlinSESM}
\sigma =  v_1+h_1, \qquad
\PhiM = {\textstyle\frac{1}{\sqrt{2}}} (v_2 + h_2) U(\bfzeta),
\end{align}
with vev parameters $v_1$ and $v_2$, and
$U(\bfzeta)$ denoting the same unitary would-be Goldstone-boson matrix as in the SM,
where $v=v_2$.
The fields $(h_1,h_2)$ are rotated
into a field basis of $h$ and $H$ corresponding to mass eigenstates,
\begin{align}\label{eq:RalphaSESM}
\begin{pmatrix} h_{1} \\ h_{2} \end{pmatrix}
= R(\alpha) \begin{pmatrix} H \\ h \end{pmatrix},
\qquad
R(\alpha) = \begin{pmatrix} \cos\alpha & -\sin\alpha\\ \sin\alpha & ~\cos\alpha \end{pmatrix},
\end{align}
where $\alpha$ is a real-valued mixing angle 
which is determined by the parameters of the Higgs potential.
In analogy to the SM case, the one-point functions
$\Gamma^h_{\mathrm{nl}}$ and $\Gamma^H_{\mathrm{nl}}$ of the fields
$h$ and $H$ are gauge independent in the non-linear representation.

\myparagraph{Two-Higgs-Doublet Model (THDM)}

The THDM comprises two complex Higgs doublets $\PhiM_n$ ($n=1,2$),
which together contain eight real d.o.f.s. 
Denoting the vev parameters of $\PhiM_n$ by $v_n$, we parametrize
the $2\times2$ matrix fields according to
\looseness-1
\begin{align}\label{eq:Phinonlin}
\PhiM_n = {\textstyle\frac{1}{\sqrt{2}}} \, U(\bfzeta) \, 
\bigl[(v_n + h_n) \id +\ri c_{n}\sigma_j \rho_j \bigr],
\end{align}
with the unitary would-be Goldstone-boson matrix $U(\bfzeta)$ as in the SM with 
\begin{align}\label{eq:Coeffs}
  v \equiv  \sqrt{v_1^2 + v_2^2}, \qquad
  c_{1} = -\frac{v_2}{v} \equiv -\sin\beta, \qquad
  c_{2} =  \frac{v_1}{v} \equiv  \cos\beta.
\end{align}
The gauge-invariant fields $h_1$ and $h_2$
are CP even (under appropriate assumptions on the Higgs potential)
and are rotated into the fields $h$ and $H$ as in (\ref{eq:RalphaSESM})
corresponding to neutral CP-even mass eigenstates~h and H.
The three remaining fields $\rho_j$ describe a neutral CP-odd
Higgs boson $\PA_0$ via the gauge-invariant field $\rho_3$ and
two charged Higgs bosons $\PH^\pm$ via the fields
$\rho^\pm=(\rho_2\pm\ri\rho_1)/\sqrt{2}$. 
The proper normalization of these fields fixes the constants $c_n$ as
above. The two Higgs mixing angles $\alpha$ and $\beta$ are frequently 
taken as basic input parameters of the THDM.
The Higgs Lagrangian of the THDM is given by
\begin{align}
{\cal L}_{\text{H,THDM}} &{}=
{\textstyle\frac{1}{2}}\text{tr}\left[(D_\mu \PhiM_1)^\dagger(D^\mu \PhiM_1)\right]
+{\textstyle\frac{1}{2}}\text{tr}\left[(D_\mu \PhiM_2)^\dagger(D^\mu \PhiM_2)\right]
- V_{\text{THDM}},
\end{align}
in which the kinetic term is again non-polynomial in the 
would-be Goldstone-boson fields $\zeta_j$ but polynomial in the 
fields $h_n$ and $\rho_j$ corresponding to physical Higgs bosons.
In the non-linear Higgs representation the gauge-invariant Higgs potential
can be written as
\begin{align}
V_{\text{THDM}}={}&
{\textstyle\frac{1}{2}}{m_{11}^2}\mathrm{tr}\bigl[\PhiM_{1}^{\dagger}\PhiM_{1}\bigr]
+{\textstyle\frac{1}{2}}{m_{22}^2}\mathrm{tr}\bigl[\PhiM_{2}^{\dagger}\PhiM_{2}\bigr]
-{m_{12}^2}\mathrm{tr}\bigl[\PhiM_{1}^{\dagger}\PhiM_{2}\bigr]
\notag\\
&{}+{\textstyle\frac{1}{8}}{\lambda_{1}}\big(\mathrm{tr}\bigl[\PhiM_{1}^{\dagger}\PhiM_{1}\bigr]\big)^2
+{\textstyle\frac{1}{8}}{\lambda_{2}}\big(\mathrm{tr}\bigl[\PhiM_{2}^{\dagger}\PhiM_{2}\bigr]\big)^2
+{\textstyle\frac{1}{4}}{\lambda_{3}}\mathrm{tr}\bigl[\PhiM_{1}^{\dagger}\PhiM_{1}\bigr]\mathrm{tr}\bigl[\PhiM_{2}^{\dagger}\PhiM_{2}\bigr]
\notag\\
&{}
+\lambda_{4}\mathrm{tr}\bigl[\PhiM_{1}^{\dagger}\PhiM_{2}\Omega_+\bigr]\mathrm{tr}\bigl[\PhiM_{1}^{\dagger}\PhiM_{2}\Omega_-\bigr]
+{\textstyle\frac{1}{2}}{\lambda_{5}}\left[\big(\mathrm{tr}\bigl[\PhiM_{1}^{\dagger}\PhiM_{2} \Omega_+ \bigr]\big)^2
+\big(\mathrm{tr}\bigl[\PhiM_{1}^{\dagger}\PhiM_{2}\Omega_-\bigr]\big)^2\right]
\end{align}
with the two-dimensional projection operators $\Omega_\pm = \frac{1}{2}(1\pm \sigma_3)$,
which select the original Higgs doublet $\Phi_n$ or its charge conjugate $\Phi^{\mathrm{c}}_n$
from the matrix field $\PhiM_n$.
Obviously, the unitary Goldstone-boson matrix $U(\bfzeta)$ again drops out in 
$V_{\text{THDM}}$.
To avoid flavour-changing neutral currents at tree level,
we assume the $\mathbb{Z}_2$ symmetry $\Phi_1\to-\Phi_1$ and
$\Phi_2\to\Phi_2$ that is only softly broken by the $m_{12}^2$ term
in $V_{\text{THDM}}$.
Moreover, all couplings in $V_{\text{THDM}}$ are assumed 
to be real in order to conserve CP.
Owing to the gauge invariance of the fields $h$ and $H$
the one-point functions
$\Gamma^h_{\mathrm{nl}}$ and $\Gamma^H_{\mathrm{nl}}$ 
are again gauge independent in the non-linear representation.

\mysection{The GIVS in the SM}

Before formulating the GIVS as introduced in Ref.~\cite{Dittmaier:2022maf}
for the SM, we briefly sketch the FJTS and PRTS for treating tadpoles
in the course of renormalizing the theory.
The full renormalization procedure of the SM
can be found in Ref.~\cite{Denner:2019vbn}.
Renormalization starts with the transformation 
that replaces the original {\it bare} quantities in terms of
{\it renormalized} ones and renormalization constants.
We mark bare parameters by subscripts ``0'' and bare
fields by subscripts ``B''.
\looseness-1

In the following, we denote the physical Higgs-boson field
of the usually adopted linear Higgs representation by 
$v+\eta(x)$.
One-particle-irreducible Green function $\Gamma^{\dots}$, 
so-called {\it vertex functions}, 
involve tadpole contributions if the splitting $v_0+\eta_\bare(x)$ 
does not provide an expansion of the effective Higgs potential about its true
minimum (see for instance App.~C of Ref.~\cite{Denner:2018opp}).  
Technically, it is desirable to eliminate such tadpole contributions
by appropriate parameter and field definitions.
Choosing $v_0$ such that $v_0^2=4\mu_{2,0}^{2}/\lambda_{2,0}$ at least to leading order (LO)
avoids tadpole contributions at tree level.
In higher orders, the explicit (unrenormalized)
tadpole function $\Gamma^\eta$ 
can be cancelled upon generating a tadpole coun\-ter\-term 
$\delta t\, \eta$ in the coun\-ter\-term Lagrangian $\delta {\cal L}$.
This is achieved by a 
tadpole renormalization condition for the renormalized 
one-point function $\Gamma_{\ren}^\eta$ (in momentum space)
of the physical Higgs field,
\begin{align}
\Gamma_{\ren}^\eta = \Gamma^\eta + \delta t \overset{!}{=} 0
\quad\Rightarrow\quad
\delta t = - \Gamma^\eta.
\label{eq:tadCT}
\end{align}
Note that $\Gamma^\eta$ is a gauge-dependent quantity in contrast to its
counterpart $\Gamma^h_{\mathrm{nl}}$ in the non-linear representation.
The tadpole coun\-ter\-term $\delta t\, \eta$ is generated 
in the Lagrangian
by appropriately choosing $v_0$
and, if needed, by a further redefinition of the bare Higgs field $\eta_\bare$.
Inserting the field decomposition $v_0+\eta_\bare(x)$ into the bare SM
Lagrangian ${\cal L}$, produces a term $t_0\, \eta$ in ${\cal L}$ with
\begin{align}
t_0 = {\textstyle\frac{1}{4}} v_0 (4\mu_{2,0}^2-\lambda_{2,0}v_0^2)
\label{eq:t0}
\end{align}
at the one-loop level,
where $t_0$ can be viewed as {\it bare tadpole constant}.
The tadpole schemes described below impose different conditions on $t_0$,
partially accompanied by appropriate field redefinitions of $\eta_\bare$,
in order to generate the desired tadpole coun\-ter\-term $\delta t h$ in the
coun\-ter\-term Lagrangian $\delta {\cal L}$.

In the {\it FJTS} the bare tadpole constant is consistently set to zero,
$t_0 = 0$, 
so that no tadpole coun\-ter\-term is introduced via parameter redefinitions.
Instead, the tadpole coun\-ter\-term is introduced by an additional field
redefinition
\begin{align}
\eta_B(x) &{} \;\to\; \eta_B(x) + \Delta v^\FJTS, \qquad 
\Delta v^\FJTS = -\frac{\delta t^\FJTS}{\MH^2}= \frac{\Gamma^\eta}{\MH^2},
\label{eq:etashiftFJTS}
\end{align}
in the bare Lagrangian. 
The field shift (\ref{eq:etashiftFJTS})
distributes tadpole renormalization constants to many
coun\-ter\-terms in $\delta {\cal L}$
(see, e.g., App.~A of Ref.~\cite{Denner:2019vbn}).
Since the field shift (\ref{eq:etashiftFJTS}) is a mere
reparametrization of the functional integral over the Higgs field, 
it does not alter predictions for observables.
Omitting the field shift would mean that
explicit tadpole diagrams had to be included, 
but the result would still be the same as in the FJTS.
In the FJTS, tadpole contributions correct for the fact that the effective 
Higgs potential is not expanded about the location of its 
minimum,
but about the minimum of the potential in lowest order,
which in the course of renormalization receives
further corrections.
For this reason, renormalization constants to mass parameters
receive tadpole corrections in the FJTS, which are rather large
by experience. 
In OS renormalization schemes these corrections
cancel in predictions,
but in other renormalization schemes such as $\MSbar$ schemes this cancellation
is only partial, and large corrections typically remain.
On the positive side, the FJTS respects gauge invariance, i.e.\ 
the gauge independence of the parametrization of an observable in terms of
$v_0$ and the other bare parameters 
carries over to the renormalized version of these parameters
if the corresponding renormalization constants do not introduce 
gauge dependences, which is for instance the case in OS and $\MSbar$ schemes in the FJTS.

The idea behind the {\it PRTS} is to achieve an expansion of the 
Higgs field about the true minimum of the renormalized effective Higgs potential 
(as obtained from the effective action after renormalization)
by appropriate relations among the parameters of the theory.
To this end, the bare parameter $v_0=v + \delta v$
is renormalized in such a way that the renormalized parameter~$v$
is fixed by the renormalized W-boson mass $\MW$ and and the
SU(2) gauge coupling $g_2=e/\sw$
according to $v = 2\MW/g_2$,
where $\sw=\sin\theta_\rw$ is the sinus of the weak mixing angle $\theta_\rw$
and $e$ the electric unit charge.
The corresponding renormalization constant $\delta v$
is directly fixed by the renormalization conditions on
$e$, $\MW$, and $\sw^2={1-\MW^2/\MZ^2}$.
In order to guarantee the compensation of all tadpole contributions after 
renormalization, the bare tadpole constant 
$t_0 = t^\PRTS + \delta t^\PRTS$
is split into a renormalized value $t^\PRTS$ and a corresponding
renormalization constant $\delta t^\PRTS$
and we demand $t^\PRTS=0$, so that
\begin{align}
\delta t^\PRTS= v_0(\mu_{2,0}^2-{\textstyle\frac{1}{4}}\lambda_{2,0} v_0^2)
= {v}(\mu_{2,0}^2-{\textstyle\frac{1}{4}}\lambda_{2,0} v^2
-{\textstyle\frac{1}{2}}\lambda_{2,0} v\delta v ),
\label{eq:dtPRTS}
\end{align}
where the second equality holds in one-loop approximation.
Since the renormalized parameter~$v$, which is directly fixed by measurements, and 
the original bare parameters $\mu_{2,0}^2$ and $\lambda_{2,0}$ are gauge independent,
the gauge dependence of $\delta t^\PRTS$ goes over to $\delta v$,
where it shows up as gauge dependence in the mass renormalization constant $\delta\MW^2$.
Trading the two bare parameters $\mu_{2,0}^2$ and $\lambda_{2,0}$ of the Higgs sector
for $v_0$ and $M_{\PH,0}$, the PRTS tadpole counterterms can be generated
by the replacements
\begin{align}\label{eq:dtgenPRTS}
\lambda_{2,0}\;\to\;\lambda_{2,0}+\frac{2\delta t^\PRTS}{v^3}, \qquad
\mu_{2,0}^2\;\to\;\mu_{2,0}^2+\frac{3\delta t^\PRTS}{2v}
\end{align}
in the bare Lagrangian with $t_0=0$, i.e.\ 
several vertex coun\-ter\-terms receive
contributions from $\delta t^\PRTS$ (see, e.g., 
App.~A of Ref.~\cite{Denner:2019vbn}).
If $\MSbar$-renormalized mass parameters are used as input,
the gauge dependence of $\delta t^\PRTS$ enters the parametrization of
observables in the step where $\mu_{2,0}^2$ and $\lambda_{2,0}$ are traded for
$v_0$ and $M_{\PH,0}$.
Note that this flaw of introducing gauge dependences does not invalidate 
the applicability of the PRTS.
On the positive side, the PRTS has the practical advantage over the FJTS
that contributions to mass renormalization constants are much smaller,
which, in particular, implies that conversions of renormalized 
mass parameters between OS
and $\MSbar$ renormalization schemes are typically
much smaller in the PRTS as compared to the FJTS.

The {\it GIVS} aims to
unify the benefits of the FJTS and the PRTS:
the gauge-invariance property of the former and the perturbative stability
of the latter.
To avoid potentially large corrections induced by tadpole loops
as inherent in the FJTS, the vev of the Higgs field is tied
to the true minimum of the effective Higgs potential.
Gauge dependences are avoided by switching to the non-linear Higgs representation
(\ref{eq:PhinonlinSM})
where the condition $v_0=v+\delta v$ applies to the gauge-invariant
component $v_0+h_\bare(x)$.
In detail, in the non-linear Higgs representation the GIVS is identical to
the PRTS described above, i.e.\ the tadpole renormalization constant is fixed
according to
\begin{align}
\delta t^\GIVS_{\mathrm{nl}} = 
\delta t^\PRTS_{\mathrm{nl}} = -\Gamma^h_{\mathrm{nl}},
\end{align}
which is gauge independent as pointed out in the previous section.
However, actual next-to-leading-order (NLO)
calculations are typically carried out in the linear
representation, and simply taking $\delta t^\PRTS_{\mathrm{nl}}$ as
tadpole renormalization constant there does not fully cancel explicit
tadpole loops. 
In order to fix this, we calculate $\delta t^\GIVS$ in the linear
representations from 
$\delta t^\GIVS_{\mathrm{nl}}$
plus an extra term to restore
$\delta t = - \Gamma^\eta$ as demanded in (\ref{eq:tadCT}):
\begin{align}
\delta t^\GIVS = \delta t^\GIVS_1 + \delta t^\GIVS_2, \qquad
\delta t^\GIVS_1 = -\Gamma^h_{\mathrm{nl}}, \qquad
\delta t^\GIVS_2 =  \Gamma^h_{\mathrm{nl}} -\Gamma^\eta.
\end{align}
The gauge-independent part $\delta t^\GIVS_1$ occurs in PRTS-like tadpole
contributions to counterterm which are generated as in (\ref{eq:dtgenPRTS})
with $\delta t^\GIVS_1$ playing the role of $\delta t^\PRTS$.
This part absorbs potentially large corrections to the location of
the minimum in the effective Higgs potential into renormalized input
parameters.
The gauge-dependent part $\delta t^\GIVS_2$ occurs in FJTS-like tadpole
contributions to counterterm which are generated by the field shift
\begin{align}
\eta_B(x) \;\to\; \eta_B(x) + \Delta v^\GIVS, \qquad
\Delta v^\GIVS = -\frac{\delta t^\GIVS_2}{\MH^2}
\end{align}
analogous to (\ref{eq:etashiftFJTS}).
In summary, knowing
the tadpole coun\-ter\-terms of the FJTS and the PRTS, as
e.g., given in the SM Feynman rules of App.~A of Ref.~\cite{Denner:2019vbn},
the generation of the one-loop GIVS tadpole coun\-ter\-terms is easily
accomplished by the substitutions
\begin{align}
\delta t^{\PRTS} \;\to\; \delta t^\GIVS_1, \qquad
\delta t^{\FJTS} \;\to\; \delta t^\GIVS_2.
\label{eq:dtGIVSsubst}
\end{align}

In Ref.~\cite{Dittmaier:2022maf} we have illustrated the perturbative
stability of the GIVS by considering the conversion of OS-renormalized
masses of SM particles to $\MSbar$ masses with the various tadpole schemes.
Table~\ref{tab:Mdiff} shows the corresponding mass shifts
induced by NLO EW corrections for the heaviest particles in the SM.
\begin{table}
\centerline{\small
\setlength{\arraycolsep}{.4em}
$\begin{array}{|c|c|c|c|c||c|c|c|c|c|}
\hline
& M^\OS[\mathrm{GeV}]
& \multicolumn{3}{c||}{\vphantom{\rule{0em}{1.2em}}\Delta M^{\MSbar{-}\OS}_{\EW}[\mathrm{GeV}]}
& & M^\OS[\mathrm{GeV}]
& \multicolumn{3}{c|}{\vphantom{\rule{0em}{1.2em}}\Delta M^{\MSbar{-}\OS}_{\EW}[\mathrm{GeV}]}
\\[.2em]
& & \FJTS & \PRTS & \GIVS & & & \FJTS & \PRTS & \GIVS
\\
\hline
\mbox{W boson} & 80.379  & -2.22 & 0.82 & 0.74
& \mbox{top quark} & 172.4 & 10.75 & 0.99 & 0.54\\
\mbox{Z boson} & 91.1876 & -0.77  & 1.25 & 1.14
& \mbox{bottom quark} & 4.93 & -1.79 & 0.10 & 0.13\\
\mbox{Higgs boson} & 125.1 & 6.34 & 3.16 & 2.80 
& \mbox{$\tau$ lepton} & 1.77686 & -0.93 & -0.028 & -0.015\\
\hline
\end{array}$
}
\caption{On-shell masses $M^\OS$ of the heaviest SM particles and differences
\mbox{$\Delta M^{\MSbar{-}\OS}_{\EW}$} between the \MSbar mass
$\overline{M}(\mu)$ and $M^\OS$
induced by NLO EW corrections using the FJTS, PRTS, or GIVS
at the renormalization scale $\mu=M^\OS$
(taken from Ref.~\cite{Dittmaier:2022maf}).}
\label{tab:Mdiff}
\end{table}
The masses entering in $\Delta M^{\MSbar{-}\OS}_{\EW}$ are chosen according to the OS mass 
values given in Tab.~\ref{tab:Mdiff}, and 
for the PRTS the 't~Hooft--Feynman gauge is chosen.

The values obtained in the PRTS and 
the GIVS are of comparable size while in general the FJTS leads to larger differences 
between the OS and the \MSbar masses. 
As emphasized in the literature~\cite{Jegerlehner:2012kn, Kniehl:2015nwa, Kataev:2022dua}
for the top quark before,
the FJTS shift $\Delta m^{\MSbar{-}\OS}_{\Pt,\EW}=10.75\GeV$ in the conversion of fermion masses 
is much larger than the typical size of EW corrections of the percent level. 
For the lighter fermions $\Pb$ and $\tau$, the relative corrections
$\Delta M^{\MSbar{-}\OS}_{\EW}/M^\OS$ are even larger than for the top quark in the FJTS,
reaching up to $\sim50\%$, while the shifts in the PRTS and GIVS remain 
all moderate.
Despite these large corrections, the FJTS often is
favoured in the literature in this
context, since it leads to a gauge-independent result in contrast to the PRTS. 
Note, however, that these large EW one-loop corrections entail an enhancement of the
theoretical uncertainties due to missing higher-order corrections.
The GIVS, on the other hand, provides gauge-independent mass shifts that are
moderate and, thus, leads to smaller EW theory uncertainties, when those
uncertainties are estimated by the propagation of the known corrections to higher
order as typically done.

\mysection{The GIVS in extended Higgs sectors}

The generalization of the GIVS from the SM to
extended Higgs sectors is fully straightforward.
The tadpole renormalization constant 
$\delta t_{h_n}=-\Gamma^{h_n}$ of any 
Higgs field $v_n+h_n(x)$ that can acquire a vev $v_n$ 
is generated in the GIVS from two parts as in (\ref{eq:dtGIVSsubst}).
The gauge-independent PRTS-like part
$\delta t^\GIVS_{h_n,1}=-\Gamma^{h_n}_{\mathrm{nl}}$
is calculated in the non-linear Higgs representation,
and the gauge-dependent FJTS-like part
$\delta t^\GIVS_{h_n,2}=\Gamma^{h_n}_{\mathrm{nl}}-\Gamma^{h_n}$
accounts for the remaining contributions.
For the SESM and THDM, the application of the GIVS 
and its impact on the $\MSbar$ renormalization of the Higgs mixing angles
are described in Ref.~\cite{Dittmaier:2022ivi} in detail.
For the $\MSbar$ renormalization of the THDM mixing angle $\beta$ we find that
the PRTS in any $R_\xi$ gauge coincides with the GIVS, a fact that 
puts existing PRTS results on a gauge-independent basis when reinterpreted
as GIVS results. We conjecture that this feature of the $\MSbar$ renormalization
of $\beta$ also carries over to supersymmetric theories.

The renormalization of Higgs mixing angles as well as the
weaknesses and strengths of various schemes
was discussed for the SESM and THDM already in 
Refs.~\cite{Krause:2016oke,Denner:2016etu,Altenkamp:2017ldc,%
Altenkamp:2018bcs,Denner:2018opp}.
In particular, Ref.~\cite{Denner:2018opp} 
analyzes existing and newly suggested renormalization
schemes wrt.\ the following criteria:
(i)~gauge independence, 
(ii)~symmetry wrt.\ mixing degrees of freedom,
(iii)~perturbative stability, and
(iv)~smoothness for degenerate masses or extreme mixing angles.
While the suggested schemes based on field-theoretical symmetries
or on OS conditions widely meet these requirements,
$\MSbar$ renormalization with FJTS tadpole treatment turned out to
be particularly prone to
perturbative instabilities in specific parameter
regions (large or degenerate Higgs masses, extreme mixing angles).
These features were demonstrated in a comprehensive discussion
of NLO predictions for various Higgs-boson production and decay 
processes in Ref.~\cite{Denner:2018opp}, extending the earlier
discussions of Refs.~\cite{Krause:2016oke,Denner:2016etu,Altenkamp:2017ldc,%
Altenkamp:2018bcs}.

The results of Ref.~\cite{Dittmaier:2022ivi} 
for the Higgs-boson decays $\Ph/\PH\to\PW\PW/\PZ\PZ\to4f$
in the SESM and THDM demonstrate that
$\MSbar$ renormalization with GIVS tadpole treatment mitigates perturbative
instabilities significantly and produces gauge-independent results very close to the 
gauge-dependent PRTS.
Table~\ref{tab:THDM-H24f} shows some results for the decays $\Ph\to4f$ 
in two THDM scenarios, which illustrate that the scale uncertainty of the
$\MSbar$ FJTS results is not always reduced in the transition from LO to NLO.
It should also be mentioned that all $\MSbar$ variants run
into problems with perturbative stability in extreme parameter regions; 
for instance, the $\MSbar$ schemes do not give reliable results for the
$\PH\to4f$ decays of the heavy Higgs boson~H in the THDM.
\begin{table}[b]
\centerline{\renewcommand{\arraystretch}{1.25} \small \tabcolsep 2pt
\begin{tabular}{|c|c||l|l|l|l|}
\hline
 & & \multicolumn{2}{c|}{A1} & \multicolumn{2}{c|}{A2} 
\\
Ren.\ scheme & tadpoles &
\multicolumn{1}{c|}{LO} & \multicolumn{1}{c|}{NLO}  & \multicolumn{1}{c|}{LO} & \multicolumn{1}{c|}{NLO}  
\\
\hline
\hline
OS12\,($\alpha,\beta$) & &
$0.89832(3)$ & $0.96194(7)_{+0.1\%}^{-0.1\%}$ &
$0.87110(3)$ & $0.92947(7)_{+0.1\%}^{-0.2\%}$
\\
\hline
\MSbar($\alpha,\beta$) & FJTS &
$0.89996(3)_{-7.4\%}^{+0.7\%}$ & $0.96283(7)_{-0.2\%}^{+0.8\%}$ &
$0.88508(3)_{-10.0\%}^{+2.2\%}$ & $0.93604(7)_{-11.0\%}^{+3.1\%}$ 
\\
\hline
\MSbar($\alpha,\beta$) & PRTS &
$0.89035(3)_{+0.9\%}^{-2.8\%}$ & $0.96103(7)_{+0.4\%}^{+1.2\%}$ &
$0.86130(3)_{+2.3\%}^{-6.1\%}$ & $0.92784(7)_{+1.3\%}^{+1.3\%}$ 
\\
\hline
\MSbar($\alpha,\beta$) & GIVS &
$0.89082(3)_{+0.9\%}^{-2.7\%}$ & $0.96106(7)_{+0.5\%}^{+1.2\%}$ &
$0.86249(3)_{+2.3\%}^{-5.8\%}$ & $0.92808(7)_{+1.3\%}^{+1.3\%}$ 
\\
\hline
\MSbar($\lambda_3,\beta$) & FJTS &
$0.89246(3)_{+1.6\%}^{-15.1\%}$ & $0.96108(7)_{+1.9\%}^{+17.3\%}$ &
$0.85590(3)_{+5.5\%}^{-29.8\%}$ & $0.92723(7)_{+2.8\%}^{+18.3\%}$ 
\\
\hline
\MSbar($\lambda_3,\beta$) & PRTS/GIVS &
$0.89156(3)_{+1.7\%}^{-8.4\%}$ & $0.96111(7)_{+2.1\%}^{+3.8\%}$ &
$0.85841(3)_{+5.0\%}^{-12.7\%}$ & $0.92729(7)_{+2.6\%}^{+4.6\%}$ 
\\
\hline
\end{tabular}
}
\caption{LO and NLO decay widths $\Gamma^{\Ph\to4f}$[MeV]
of the light CP-even Higgs boson $\Ph$ for THDM 
scenarios A1 and A2 in different renormalization schemes,
with the on-shell scheme OS12 as input scheme (and full conversion
of the input parameters into the other schemes).
The scale variation (given in percent) corresponds to the 
scales $\mu=\mu_0/2$ and $\mu=2\mu_0$, where 
$\mu_0$ is the average Higgs mass
(for all details, see Ref.~\cite{Dittmaier:2022ivi}).}
\label{tab:THDM-H24f}
\end{table}

\section{Conclusions}

Recently we have proposed a new scheme for a gauge-invariant 
treatment of tadpole corrections in spontaneously broken gauge theories called 
{\it Gauge-Invariant Vacuum expectation value Scheme (GIVS)}.
The GIVS is a hybrid scheme of the two most commonly used tadpole
schemes, in which tadpole counterterms are introduced either via
parameter or via field renormalization transformations.
In the construction of tadpole counterterms, the GIVS makes use of non-linear
representations of Higgs sectors, in which fields that can develop
non-vanishing vacuum expectation values are gauge invariant,
but actual loop calculations are performed in the GIVS as usual.
In contrast to previously used tadpole schemes, the GIVS unifies the 
properties of gauge independence and perturbative stability. 
We have illustrated these virtues by discussing
the conversion of
on-shell-renormalized to $\MSbar$-renormalized masses in the SM
and the $\MSbar$ renormalization of Higgs mixing angles in two
frequently considered models with extended Higgs sectors---a 
singlet Higgs extension of the SM and
the Two-Higgs-Doublet Model.
We expect that the generalization of the GIVS beyond the one-loop level
works without problems.


\begin{thebibliography}{99}
\setlength{\itemsep}{0.5em}

\bibitem{Bohm:1986rj}
M.~B{\"o}hm, H.~Spiesberger, and W.~Hollik, 
{\em Fortsch. Phys.} {\bf 34} (1986) 687--751.

\bibitem{Denner:1991kt}
A.~Denner, 
  {\em Fortsch. Phys.} {\bf 41} (1993) 307--420,
  [\href{http://arxiv.org/abs/0709.1075}{{\tt arXiv:0709.1075}}].

\bibitem{Fleischer:1980ub}
J.~Fleischer and F.~Jegerlehner, 
{\em Phys. Rev.} {\bf D23} (1981) 2001--2026.

\bibitem{Actis:2006ra}
S.~Actis, et~al.,
{\em Nucl.  Phys. B} {\bf 777} (2007) 1--34,
  [\href{http://arxiv.org/abs/hep-ph/0612122}{{\tt hep-ph/0612122}}].

\bibitem{Dittmaier:2022maf}
S.~Dittmaier and H.~Rzehak, 
{\em JHEP} {\bf 05} (2022) 125,
[\href{http://arxiv.org/abs/2203.07236}{{\tt arXiv:2203.07236}}].

\bibitem{Dittmaier:2022ivi}
S.~Dittmaier and H.~Rzehak, 
  \href{http://arxiv.org/abs/2206.01479}{{\tt arXiv:2206.01479}}.

\bibitem{Krause:2016oke}
M.~Krause, et~al.,
{\em JHEP} {\bf 09} (2016) 143, [\href{http://arxiv.org/abs/1605.04853}{{\tt
  arXiv:1605.04853}}].

\bibitem{Denner:2016etu}
A.~Denner, et~al.,
{\em JHEP} {\bf 09} (2016)
  115, [\href{http://arxiv.org/abs/1607.07352}{{\tt arXiv:1607.07352}}].

\bibitem{Denner:2019vbn}
A.~Denner and S.~Dittmaier, 
{\em Phys. Rept.} {\bf 864} (2020) 1--163,
[\href{http://arxiv.org/abs/1912.06823}{{\tt arXiv:1912.06823}}].

\bibitem{Dudenas:2020ggt}
V.~D\={u}d\.{e}nas and M.~L\"oschner,
Phys. Rev. D \textbf{103} (2021) no.7, 076010
[arXiv:2010.15076].

\bibitem{Gambino:1999ai}
P.~Gambino and P.~A. Grassi, 
{\em Phys. Rev. D} {\bf 62} (2000) 076002,
  [\href{http://arxiv.org/abs/hep-ph/9907254}{{\tt hep-ph/9907254}}].

\bibitem{Denner:2018opp}
A.~Denner, S.~Dittmaier, and J.-N. Lang, 
{\em JHEP} {\bf 11} (2018) 104,
  [\href{http://arxiv.org/abs/1808.03466}{{\tt arXiv:1808.03466}}].

\bibitem{Jegerlehner:2012kn}
F.~Jegerlehner, M.~Y. Kalmykov, and B.~A. Kniehl, 
{\em Phys. Lett. B} {\bf 722} (2013) 123--129,
  [\href{http://arxiv.org/abs/1212.4319}{{\tt arXiv:1212.4319}}].

\bibitem{Kniehl:2015nwa}
B.~A. Kniehl, A.~F. Pikelner, and O.~L. Veretin, 
{\em Nucl. Phys. B} {\bf 896} (2015) 19--51, [\href{http://arxiv.org/abs/1503.02138}{{\tt
  arXiv:1503.02138}}].

\bibitem{Kataev:2022dua}
A.~L. Kataev and V.~S. Molokoedov, 
  \href{http://arxiv.org/abs/2201.12073}{{\tt arXiv:2201.12073}}.

\bibitem{Altenkamp:2017ldc}
L.~Altenkamp, S.~Dittmaier, and H.~Rzehak, 
   {\em JHEP} {\bf 09} (2017) 134, [\href{http://arxiv.org/abs/1704.02645}{{\tt
  arXiv:1704.02645}}].

\bibitem{Altenkamp:2018bcs}
L.~Altenkamp, M.~Boggia, and S.~Dittmaier, 
{\em JHEP} {\bf 04} (2018) 062,
[\href{http://arxiv.org/abs/1801.07291}{{\tt arXiv:1801.07291}}].

\end{thebibliography}
\end{document}